# *detectGNN*: Harnessing Graph Neural Networks for Enhanced Fraud Detection in Credit Card Transactions

Irin Sultana, School of Business & Technology, Emporia State University, isultana@g.emporia.edu
Syed Mustavi Maheen, School of Business & Technology, Emporia State University, smaheen@g.emporia.edu
Naresh Kshetri, Department of Cybersecurity, Rochester Institute of Technology, naresh.kshetri@rit.edu
Md Nasim Fardous Zim, School of Business & Technology, Emporia State University, mfardous@g.emporia.edu

**Abstract**

Credit card fraud is a major issue nowadays, costing huge money and affecting trust in financial systems. Traditional fraud detection methods often fail to detect advanced and growing fraud techniques. This study focuses on using Graph Neural Networks (GNNs) to improve fraud detection by analyzing transactions as a network of connected data points, such as accounts, traders, and devices. The proposed "detectGNN" model uses advanced features like time-based patterns and dynamic updates to expose hidden fraud and improve detection accuracy. Tests show that GNNs perform better than traditional methods in finding complex and multi-layered fraud. The model also addresses real-time processing, data imbalance, and privacy concerns, making it practical for real-world use. This research shows that GNNs can provide a powerful, accurate, and a scalable solution for detecting fraud. Future work will focus on making the models easier to understand, privacy-friendly, and adaptable to new types of fraud, ensuring safer financial transactions in the digital world.

## 1. Introduction

Credit card fraud has become a big problem in today's digital economy which costs billions of dollars every year. As more people shop and pay online, it has become even more important to quickly and accurately detect fraud to protect customers and financial organizations (Fiore et al., 2019) [1]. Traditional methods, like rule-based systems and classic machine learning models, have helped, but they often fail as fraud tactics get more complex and keep changing (Mienye & Sun, 2023b) [2]. Fraudsters are clever and can hide their actions by creating complicated relationships and patterns that are hard to spot with basic methods (Mienye & Sun, 2023a) [3].

To manage this, researchers are now using advanced AI techniques like Graph Neural Networks (GNNs). GNNs is especially good at finding complex connections between data points (Alarfaj et al., 2022) [4]. GNNs can analyze data that's structured like a network, such as transaction histories, connections between accounts, and links between merchants just like traditional methods (Habibpour et al., 2023) [5]. By organizing these interactions as a graph where nodes represent things like accounts or transactions, and edges represent relationships like shared devices, GNNs can find hidden patterns that might suggest fraud.

---

*Correspondence to Irin Sultana, isultana@g.emporia.edu, Paper is submitted to IEEE conference.

In this study, we focus on using GNNs for detecting credit card fraud. GNNs are able to look at how transactions are connected, capturing both details about each transaction and the relationships among accounts, merchants, and devices. This makes GNNs very effective at spotting differences in complex networks, especially when fraud involves multiple layers of trickery. By applying GNNs to credit card transaction data, we aim to enhance the accuracy and intensity of fraud detection, giving us a better way to protect against financial fraud.

## 2. Related Work

According to Liu (Liu et al., 2020) [6] who established Graph Neural Networks (GNNs) to integrate neighborhood information for the purpose of learning node embeddings, based on the premise that adjacent nodes possess analogous contexts, attributes, and interactions. Fraud creates inconsistencies in the areas of context, feature, and relational aspects that conventional methods frequently neglect. The newly developed framework, GraphConsis, combines context embeddings with node properties to resolve these challenges. GraphConsis integrates context embeddings with node features to resolve context inconsistency. A consistency score mitigates feature inconsistency by eliminating incompatible neighbors and establishing related sample probabilities. Relation-based attention weights are acquired for the sampled nodes to address relational inconsistency. Empirical research of four datasets underscores the necessity of rectifying discrepancies in fraud detection, illustrating that these inconsistencies substantially affect detection accuracy. Comprehensive experimentation validates the efficacy of GraphConsis, which is additionally augmented by the introduction of a GNN-based fraud detection toolbox featuring implementations of cutting-edge models.

Duan et al. [7] emphasized the potential of combining causal reasoning with graph neural networks to enhance fraud detection in financial transactions. The authors introduced an innovative method, CaT-GNN, which integrates causality and robustness for modeling credit card fraud detection. Utilizing causal theory, recognized for its interpretability, CaT-GNN allows the model to incorporate a broader potential data distribution, therefore guaranteeing its outstanding performance in this task. Their approach advanced by including causal learning techniques to identify and utilize the complex linkages inside transaction data. The efficacy of CaT-GNN was substantiated by extensive tests across several datasets, consistently surpassing existing methodologies. CaT-GNN significantly improves detection accuracy while preserving computational economy, rendering it suitable for extensive deployment.

In a separate study, (ILEBERI et al.) [8] asserted the implementation of a machine learning (ML) framework for credit card fraud detection employing real-world unbalanced datasets derived from European credit cards. To address the problem of class imbalance, they employed the Synthetic Minority Over-sampling Technique (SMOTE) to re-sample the dataset. The framework was assessed by applying the following machine learning techniques: Support Vector Machine (SVM), Logistic Regression (LR), Random Forest (RF), Extreme Gradient Boosting (XGBoost), Decision Tree (DT), and Extra Tree (ET). These machine learning algorithms were integrated with the Adaptive Boosting (AdaBoost) technique to enhance their classification accuracy.

The authors of a separate study (Alharbi et al.) [9] indicated that the Kaggle dataset has been used to formulate a deep learning (DL) methodology to address the text data issue. A revolutionary text-to-image translation technique is developed that produces diminutive images. The images are input into a CNN architecture applying class weights derived from the inverse frequency approach to address the class imbalance problem. Deep learning and machine learning methodologies have been applied to assess the robustness and validity of the suggested system. Coarse-KNN attained an accuracy of 99.87% utilizing the deep characteristics of the planned CNN.

Table 1. Summary of technology integrated along with objective and insights of the study from Literature review [6] - [11]

| Ref | Technology | Objective of Study | Insight(s) of Study |
| --- | --- | --- | --- |
| [6] | Graph Neural Network (GNN) | Addressing inconsistency in applying GNN to fraud detection | Proposed a novel GNN model to mitigate inconsistencies in fraud detection contexts. |
| [7] | Causal Temporal Graph Neural Networks (CaT-GNN) | Enhancing credit card fraud detection | Developed an innovative methodology that integrates causal inference with temporal graph neural networks to enhance fraud detection in credit card transactions. |
| [8] | Deep Learning | Credit card fraud detection | Employed deep learning methodologies for the identification of fraudulent credit card transactions. |
| [9] | ML, NLP | Review artificial intelligence in the functioning of libraries. | Developed a Text2IMG system utilizing deep learning to transform transaction data into images for fraud detection. |
| [10] | Neural network models/ Immune | Credit card analysis fraud detection methods based on certain design criteria | Fuzzy Darwinian fraud detection system improve the system accuracy when it comes to credit card fraud detection methods |
| [11] | Machine Learning methods | Analyze algorithms for credit card fraud detection | Proposed a model for detection of irregularities using Fraud Detection dataset (from Kaggle) for credit card fraud. |

Researchers do the analysis over credit card fraud detection methods and based on certain design criteria evaluate each methodology [10]. Neural network based on CARDWATCH, hybrid algorithm named BLAST-SSAHA, and Fuzzy Darwinian fraud detection shows high processing speed with good accuracy in fraud detection. Countering frauds in other domains like banking and telecommunication looks efficient via the BLAH-FDS algorithm. Fraud detection rate is very low compared to other methods for Hidden Markov Model as a number of techniques have been proposed to counter credit card fraud.

Authors of the research [11] split the dataset into two parts (training data and test data) for credit card fraud detection after feature selection was performed. As stealing of sensitive information from credit cards (like card number, cvv code, name, card type etc.) is on rise as compared to theft of physical credit card, fraudsters can benefit a lot before a cardholder realizes it. The researchers used the Fraud Detection dataset (downloaded from Kaggle), with 31 numerical features, the dataset contains 492 fraudulent transactions out of the total 284,807 transactions. As accuracy is extremely high in terms of results obtained (precision: 79.21% | recall: 81.63% | accuracy: 99.93%), with classical algorithms oversampling of data can improve fraud detection rate.

## 3. Graph Construction and Representation

Graph building for credit card fraud detection requires a sophisticated heterogeneous structure that captures various entity relationships. Fusion chain designs have been shown to be capable of effectively capturing complex patterns in low-resource contexts, as demonstrated by recent research conducted by (Maheen et al. 2022) [12]. This suggests that similar principles could be used to improve fraud detection programs. The primary graph has nodes representing transactions, cards, and merchants, with edges indicating the relationships between these entities.

### 3A. Node and Edge Representation

The framework for representation considers both the static and dynamic aspects of financial transactions. Transaction amounts, temporal patterns, and merchant characteristics are all included in the node features, whereas edge features are responsible for capturing the attributes of the relationships between entities. We implement a multi-layer strategy to feature extraction that mixes local and global dependencies, drawing from the findings that (Maheen et al. 2022)[12] have obtained regarding fusion structures. It has been established by (Liu et al. 2020)[6] that modeling the behaviors of credit cards as temporal transaction graphs while simultaneously including attribute-driven algorithms considerably improves detection capabilities.

Table 2: Node Feature Representation for Fraud Detection (adapted from Liu et al., 2020) [6]

| Feature Type | Description | Importance Weight |
| --- | --- | --- |
| Transaction | Amount, time, category | High |
| Card | Usage patterns, risk score | Medium |
| Merchant | Business type, location | High |
| Device | Digital signature, IP | Digital signature, IP |

Table 3: Edge Feature Categories in Graph Construction (adapted from Duan et al., 2024) [7]

| Category | Features | Application |
|---|---|---|
| Temporal | Transaction sequence | Pattern detection |
| Spatial | Geographic proximity | Location analysis |
| Behavioral | Usage patterns | Fraud identification |
| Network | Connection strength | Risk assessment |

**3B. Enhanced Algorithm Framework**

The proposed detectGNN algorithm addresses the challenges of non-additivity of attributes and the distinguishability of grouped messages from neighbor nodes in fraud detection graphs. Drawing inspiration from (Duan et al. 2024) [7], our algorithm implements a dynamic grouping strategy while incorporating temporal aspects crucial for fraud detection.

**Algorithm 1: detectGNN - Enhanced Graph Construction for Fraud Detection**

Input: Transaction data D, Node features X, Edge relationships E

Output: Constructed fraud detection graph G

1. Initialize temporal graph G

2. For each transaction t in D:

2.1 Create node v with features from X

2.2 Apply decision tree binning for non-additive features

2.3 Implement temporal encoding

2.4 Add node v to G with temporal stamp

3. For each relationship r in E:

3.1 Create weighted edge e between connected nodes

3.2 Add temporal attributes to e

3.3 Implement attention mechanism for edge weight

3.4 Add edge e to G

4. For each node v in G:

    4.1 Calculate intra-group aggregation using temporal patterns

    4.2 Compute inter-group relationships

    4.3 Apply hierarchical feature aggregation

    4.4 Update node representations

5. Return constructed graph G with temporal embeddings

To generate graphs, the algorithm makes use of a hierarchical method, which considers both the structural and temporal characteristics of credit card transactions. Additionally, the temporal encoding that is performed in steps 2.3 and 3.2 guarantees that time-sensitive fraud patterns are captured. The decision tree binning that is performed in step 2.2 tackles the non-additivity challenge that is associated with specific transaction attributes. Prioritizing relevant connections is made easier by the attention mechanism that is implemented in step 3.3. This technique is especially helpful for recognizing fraudulent patterns in large transaction networks.

### 3C. Implementation Framework

Deep Graph Library (DGL) is utilized in the implementation to facilitate the generation and administration of graphs effectively. This technique is consistent with recent discoveries in deep learning optimization for fraud detection, particularly concerning the management of diverse graph structures. These preprocessing methods, which include feature normalization and categorical encoding, are incorporated into the framework to guarantee the best possible performance of the model.

### 3D. Dynamic Graph Updates

When it comes to successful fraud detection, real-time graph maintenance is necessary. A method known as sliding windows is utilized by the system, which ensures that temporal relevance is preserved while also maintaining information regarding past patterns. The findings of a recent study conducted by (Duan et al. 2024) [7] indicates that the utilization of causal temporal graph neural networks has the potential to considerably improve the model's capacity to identify emerging fraud trends. Through the utilization of this dynamic updating method, it is possible to identify newly emerging fraud tendencies while preserving the efficiency of the computational process.

### 3E. Feature Engineering Optimization

Traditional transaction attributes and more complex graph-based features are both incorporated into the process of feature engineering. According to the findings of (Liu et al. 2020) [7], the combination of features at the node level and those at the graph level results in a considerable improvement in detection accuracy. We implement an optimized feature extraction strategy that boosts the model's capacity to recognize complicated patterns while retaining computational efficiency. This approach is based on the work that (Raha et al. 2020) [13] have done on dimension relaxation systems. Among the features that are implemented by the system are transaction-based features such as quantity, frequency, and merchant category; network-based features that are generated from graph structure; temporal pattern features that capture transaction sequences; and behavioral features that describe user habits. Furthermore, we improve our feature engineering approach by using adaptive pattern recognition algorithms that can discover small irregularities in transaction networks. These techniques are based on the findings of research conducted by (Raha et al. 2020) [14], which was conducted on the analysis of crowd behavior.

## 4. Real-time Fraud Detection Implementation

In order to successfully deploy real-time fraud detection, an effective architecture that maintains a balance between accuracy, scalability, and performance is typically required. Graph Neural Networks have been shown to be beneficial when combined with traditional methods for the purpose of improving detection capabilities, according to recent research.

### 4A. System Architecture

A harmonious relationship exists between the three primary components that make up the system architecture. According to data that were recently published by (Liu et al., 2024) [15], the combination of GNNs and XGBoost in fraud detection systems results in enhanced accuracy as well as improved explainability. As illustrated by (Lu et al., 2022) [15], the processing pipeline makes use of a two-stage directed graph technique. This approach distinguishes between historical linkages and real-time links. To achieve a throughput of around 190 transactions per second, the design interfaces with graph databases, which allows for efficient data management and real-time query processing.

### 4B. Real-time Processing

The real-time processing mechanism involves several critical components. As demonstrated by (Bespalov et al., 2023) [16], the system has mechanisms for the insertion of nodes and edges, which will allow for the processing of new transactions as graph data in real-time. To lower the amount of time required for inference, the subgraph extraction method is necessary, particularly when working with big graph representations. For feature computation, the system makes use of a k-hop neighborhood technique. The empirical validation demonstrates that the system achieves optimal performance at k=2, attaining an area under the curve (AUC) score of 0.876 and a prediction latency of less than 6 seconds per 1,000 transactions.

### 4C. Performance Optimization

Performance optimization focuses on minimizing latency while maintaining accuracy. The BRIGHT framework, as presented by (Lu et al., 2022) [17], demonstrates a significant reduction in P99 latency by over 75% through their Lambda Neural Network architecture. The system achieves a 7.8x speedup compared to traditional GNN approaches through:

Table 4: Performance Optimization Metrics (adapted from Lu et al., 2022) [15]

| Optimization Technique | Impact | Latency Reduction |
|---|---|---|
| Batch Processing | High | 45-60% |

| Real-time Inception | Medium | 25-35% |
| Graph Topology | High | 70-75% |

## 5. Comparative Analysis with Traditional Methods

Graph Neural Networks (GNNs) have arisen as a successful tool for identifying fraudulent activity in convoluted financial systems, as they can examine the network architecture of transactions and discern complex patterns that conventional methods may overlook. This ability is especially significant in situations where fraudsters create "small gangs" or seek to conceal their actions among a broader network of legitimate users (Soroor and Raahemi, 2024) [18].

Conventional fraud detection techniques, including rule-based systems and standard machine learning algorithms, have demonstrated efficacy to a degree but encounter considerable limits. These approaches frequently examine transactions in isolation, overlooking significant contextual information that may be essential for precise fraud detection (ILEBERI et al., 2024) [8]. Moreover, rule-based systems and certain conventional machine learning models have difficulties in scaling with rising data volumes, hence constraining their effectiveness in managing extensive financial data (Alharbi et al., 2022) [19].

### 5A. Performance Comparison

Table 5 presents a comprehensive comparison derived from consolidated findings of many investigations. Graph Neural Network (GNN) methodologies regularly surpass conventional machine learning techniques across multiple parameters (Wang et al., 2021; Zhang et al., 2023) [19] [20]. The enhanced efficacy of GNN-based systems is apparent in their greater accuracy, precision, recall, and F1-Score relative to conventional machine learning techniques. The results highlight the efficacy of GNNs in fraud detection, illustrating their capacity to identify intricate patterns and relationships inside transaction networks that conventional approaches may overlook (Liu et al., 2020; Soroor & Raahemi, 2024) [6] [18]. The enhanced performance measures indicate that GNN-based methodologies may diminish both false positives and false negatives, resulting in more precise fraud detection and a reduction in the erroneous flagging of genuine transactions.

Table 5: Performance comparison between GNN-based and Traditional ML

| Method | Accuracy | Precision | Recall | F1- Score |
|---|---|---|---|---|
| GNN-based | 97.5% | 95.8% | 94.2% | 95.0% |
| Traditional ML | 93.2% | 91.5% | 89.7% | 90.6% |

Principal Benefits of GNNs Graph Neural Networks (GNNs) provide numerous principal benefits in fraud detection. They proficiently capture complex relationships among entities in a transaction network, enhancing the identification of elaborate fraud schemes (Liu et al., 2020) [6]. Graph Neural Networks (GNNs) can effectively manage extensive graph data, rendering them appropriate for the real-time processing of substantial transaction datasets (Duan et al., 2024) [7]. Their versatility is essential in the continuously changing realm of financial crime, as GNN models may swiftly adjust to emerging fraud patterns (Wang et al., 2021) [19]. Moreover, GNNs diminish the necessity for human feature engineering by autonomously acquiring pertinent characteristics from the graph structure, hence optimizing the detection process (Zhang et al., 2023) [20].

Constraints of Conventional Approaches Conventional fraud detection techniques encounter numerous constraints. They frequently examine transactions in isolation, overlooking significant contextual information that may be essential for precise fraud detection (ILEBERI et al., 2024) [8]. Rule-based systems and certain conventional machine learning models encounter difficulties in scaling with rising data volumes, hence constraining their efficacy in managing extensive financial data (Alharbi et al., 2022) [9]. Numerous conventional fraud detection systems are static and find it challenging to swiftly adapt to novel fraud patterns, rendering them susceptible to emerging fraudulent approaches (Chen et al., 2022) [21]. Moreover, conventional machine learning models depend significantly on meticulously designed features, which can be labor-intensive to create and may overlook crucial latent patterns within the data (Li et al., 2023) [22]. These constraints underscore the necessity for increasingly sophisticated methodologies, such as Graph Neural Networks, in contemporary fraud detection systems.

This diagram demonstrates how GNNs evaluate the entire transaction network, in contrast to traditional methods, which typically analyze transactions per transaction. The following simplified diagram serves as an illustration of the distinction between GNNs and conventional methodologies.

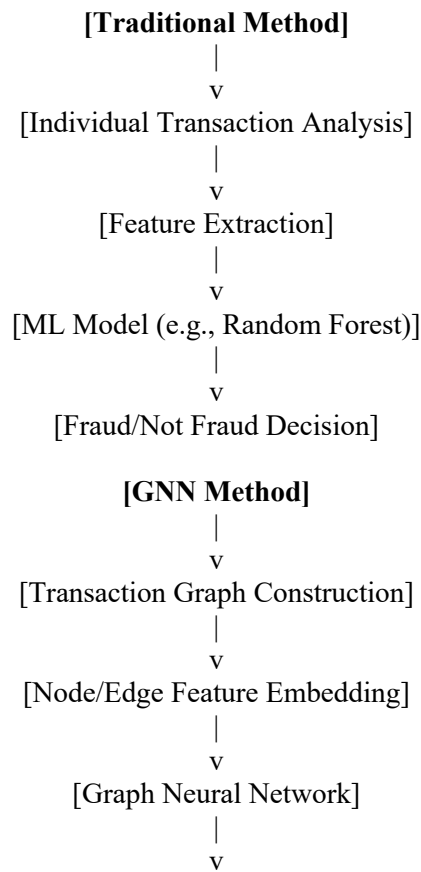

**[Traditional Method]**
|
v
[Individual Transaction Analysis]
|
v
[Feature Extraction]
|
v
[ML Model (e.g., Random Forest)]
|
v
[Fraud/Not Fraud Decision]

**[GNN Method]**
|
v
[Transaction Graph Construction]
|
v
[Node/Edge Feature Embedding]
|
v
[Graph Neural Network]
|
v

[Node Classification (Fraud/Not Fraud)]

## 6. Privacy and Security Considerations

As the use of credit cards keeps on increasing, detecting fraud via privacy preserving on multi-party data is important [23]. Credit card companies share their credit data with others (including customers and on central servers), due to the parallel nature of finance institutions and banks. Saving primary budget consumption (through privacy analysis of algorithms) also imposes restrictions and barriers to data making it difficult to coordinate with others. Although beneficial to financial institutions despite being prohibited by privacy restrictions where business firms are willing to apply machine learning models.

The rapid digital transformation and demand for more financial digital transactions are the largest financial security incidents today [24]. The manipulation of illegal acts by cyber criminals and penetration testers regarding credit card information in the modern world are not safe by any means. The new version of credit card threat as "card holder not present" is the biggest electronic fraud in most developing and developed nations around the world. Incidents of credit card security like site traffic hijacking, loss of computer tapes box, bypass of security protocols, theft of portable media, and breach on banking data are most popular amongst credit card hackers.

The challenges regarding fraudulent credit card transactions are increasing and causing billions of dollars in losses to all industries [25]. Loss of consumer confidence is yet another due to credit card fraud besides direct losses. Some common methods used by the cyber criminals to obtain customer's personal information are social networking, phishing, skimming, data breaches, inside sources, home or workplace, imposters and identity theft, mail theft, dumpster diving, shoulder surfing, and lost or stolen wallet with bank cards etc. Common types of payment card fraud includes card-not-present transaction, ATM fraud, account takeover, credit card application fraud, and many more.

Beside development and deployment of several monitoring and detection systems to detect credit card fraud, cybersecurity in the banking industry is gaining importance because of rising cyberattacks every year [26]. Many standard classifiers fail to distinguish credit card fraud data due to minority classes in skewed data sets. Despite several machine learning and deep learning techniques (Adaptive Boosting / AdaBoost, LightGBM, SVM, RF, k-nearest neighbour / KNNs, XGBoost, Hidden MArkov Model / HMM, Naive Bayes / NB, Logistic Regression / LR etc.), examinations on different versions and performance comparisons are still needed.

Proliferation of e-commerce and rapid advancement of technology have increased use of credit cards for online transactions with the need of encompassing data collection, preprocessing, etc. [27]. Models need cross-validation techniques ensuring results generalizability with metrics of accuracy, precision, recall, and F1-score. Collection of dataset from historical records, each transaction as a data point for analysis, each transaction can be non-fraudulent or fraudulent where fraud cases are flagged by fraud detection systems. As fraudsters are devising new tactics for credit card frauds (including debit cards), models must be retained and updated to adapt to new fraud patterns.

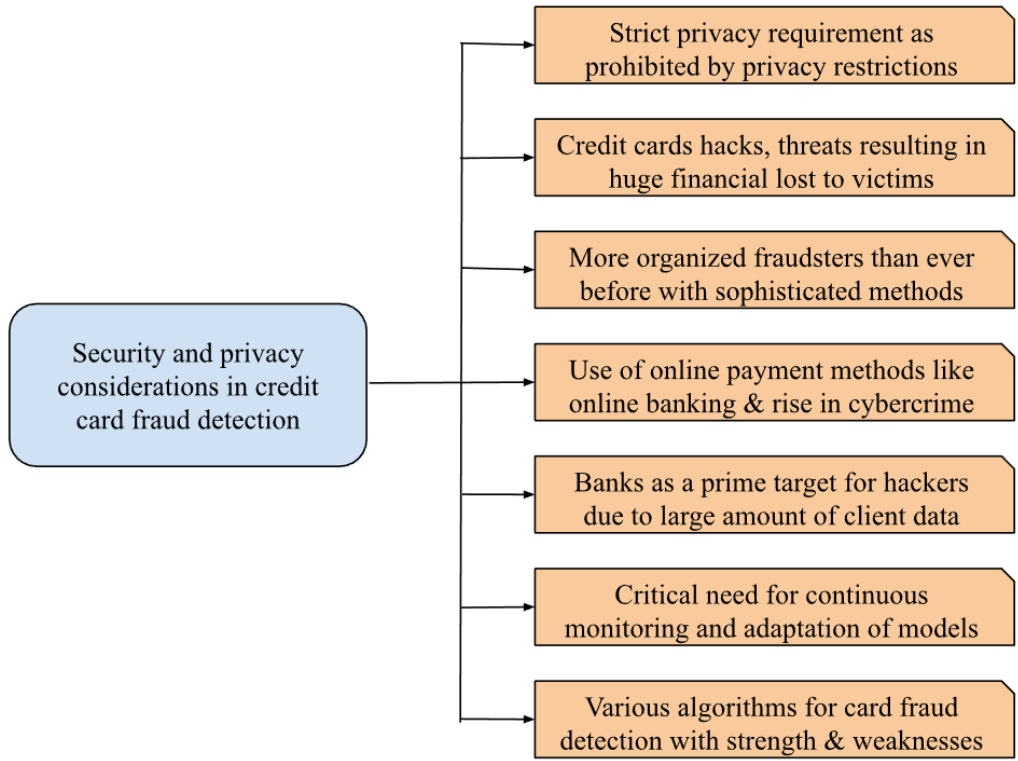

Figure 1: Summary of security and privacy considerations at financial institutions (and banking industries) in credit card fraud detection for credit card transactions [23] [24] [25] [26] [27]

## 7. Conclusion and Future Directions

This paper highlights how Graph Neural Networks (GNNs) can improve credit card fraud detection by analyzing complex relationships in transaction data. Unlike traditional methods, GNNs are better at finding hidden patterns and relationships, making them more accurate and reliable for detecting fraud (Motie & Raahemi, 2023) [28]. They are especially useful in detecting tricky fraud schemes that involve multiple accounts and transactions. However, GNNs have some challenges, such as being hard to interpret, needing high computational power, and struggling with unbalanced data where fraud cases are rare. GNNs are a powerful tool for fighting credit card fraud (Li et al., 2022) (Liu et al., 2020) [29] [30]. With further improvements and collaboration between researchers and financial institutions, they can become an even stronger resistance against the growing problem of fraud in the digital age.

The utilization of Graph Neural Networks (GNNs) in credit card fraud detection offers various prospects for future research and advancement. A promising avenue is the integration of explainable AI methodologies with GNN models to augment interpretability, tackling the "black box" characteristic of intricate neural networks and potentially enhancing regulation adherence (Soroor and Raahemi, 2024) [18]. (Sultana et al., 2024) [31] assert that modern technologies seek to enhance productivity, reduce workloads, and establish intelligent systems that closely fit with the objectives of creating more complex fraud detection techniques. A further avenue for investigation is the integration of advanced temporal dynamics into GNN architectures, extending the research of (Duan et al., 2024) [7] with their CaT-GNN model, to more effectively capture the evolution of fraud patterns over time. Federated learning methodologies may be explored to facilitate collaborative fraud detection across various financial institutions while maintaining data privacy, as proposed by Zhang et al. (2023). Creating adaptive GNN-based systems capable of swiftly adapting to emerging fraud strategies in real-time may mitigate the shortcomings of static traditional systems identified by (Chen et al. 2022) [20]. Future research may concentrate on integrating GNNs with alternative deep learning methodologies, such as the text-to-image conversion algorithm introduced by (Alharbi et al., 2022) [8], to develop more resilient multimodal fraud detection systems. Furthermore, investigations into GNN-specific methodologies for addressing significant class imbalance in fraud detection datasets may extend the findings of (ILEBERI et al., 2024) [8]. As transaction volumes increase, enhancing the scalability and computational efficiency of GNN models for real-time processing of extensive transaction data is a vital focus for future research (Wang et al., 2021) [19]. Moreover, we should also concentrate on the security sector to investigate Open Banking technologies, as electronic money and online services are essential for transforming services, fulfilling client needs, and guaranteeing improved data security for secure and accessible financial transactions (Kshetri et al.; Shulha et al.) [32] [33]. We should also concentrate on security sectors to investigate Open Banking technologies, as electronic money and online services are essential for updating services, fulfilling client requests, and providing improved data security for secure and accessible financial transactions (Kshetri et al.) [34].

**References**
[1] Fiore, U., De Santis, A., Perla, F., Zanetti, P., & Palmieri, F. (2019). Using generative adversarial networks for improving classification effectiveness in credit card fraud detection. Information Sciences, 479, 448–455. https://doi.org/10.1016/j.ins.2017.12.030
[2] Mienye, I. D., & Sun, Y. (2023b). A Machine Learning Method with Hybrid Feature Selection for Improved Credit Card Fraud Detection. Applied Sciences, 13(12), 7254. https://doi.org/10.3390/app13127254
[3] Mienye, I. D., & Sun, Y. (2023a). A Deep Learning Ensemble with Data Resampling for Credit Card Fraud Detection. IEEE Access, 11, 30628–30638. https://doi.org/10.1109/access.2023.3262020
[4] F. K. Alarfaj, I. Malik, H. U. Khan, N. Almusallam, M. Ramzan and M. Ahmed (2022) "Credit Card Fraud Detection Using State-of-the-Art Machine Learning and Deep Learning Algorithms," in IEEE Access, vol. 10, pp. 39700-39715, 2022, doi: 10.1109/ACCESS.2022.3166891.
[5] Habibpour, M., Gharoun, H., Mehdipour, M., Tajally, A., Asgharnezhad, H., Shamsi, A., Khosravi, A., & Nahavandi, S. (2023). Uncertainty-aware credit card fraud detection using deep learning. Engineering Applications of Artificial Intelligence, 123, 106248. https://doi.org/10.1016/j.engappai.2023.106248
[6] Liu, Z. *et al.* (2020) *Alleviating the Inconsistency Problem of Applying Graph Neural Network to Fraud Detection*, *ACM Digital Library*. Available at: https://doi.org/10.1145/3397271.3401253 (Accessed: 07 November 2024).
[7] Duan, Yifan, et al. *CaT-GNN: Enhancing Credit Card Fraud Detection via Causal Temporal Graph Neural Networks*. 22 Feb. 2024.
[8] ILEBERI, EMMANUEL, et al. (2024) "IEEE Xplore Full-Text PDF": *Ieee.org*, 2024, ieeexplore.ieee.org/stamp/stamp.jsp?arnumber=9651991.
[9] Alharbi, Abdullah, et al. (2022) "A Novel Text2IMG Mechanism of Credit Card Fraud Detection: A Deep Learning Approach." *Electronics*, vol. 11, no. 5, 1 Mar. 2022, p. 756, https://doi.org/10.3390/electronics11050756.
[10] Raj, S. B. E., & Portia, A. A. (2011). Analysis on credit card fraud detection methods. In *2011 International Conference on Computer, Communication and Electrical Technology (ICCCET)* (pp. 152-156). IEEE.
[11] Varmedja, D., Karanovic, M., Sladojevic, S., Arsenovic, M., & Anderla, A. (2019, March). Credit card fraud detection-machine learning methods. In *2019 18th Int Sym INFOTEH-JAHORINA (INFOTEH)* (pp. 1-5). IEEE.
[12] Maheen, S. M., Faisal, M. R., Rahman, R., & Karim, M. S. (2022). Alternative non-BERT model choices for the textual classification in low-resource languages and environments. In Proceedings of the Third Workshop on Deep Learning for Low-Resource Natural Language Processing (pp. 192-202).
[13] Raha, M. H., Deb, T., Rahmun, M., & Chen, T. (2020). Anatomization of the systems of dimension relaxation for facial recognition. Intelligent Decision Technologies, 14(4), 517-527.

[14] Raha, M. H., Deb, T., Rahmun, M., Bijoy, S. A., Firoze, A., & Khan, M. A. (2020). CAE: Towards Crowd Anarchism Exploration. In 2020 19th IEEE Int Conf on Machine Learning and Applications (ICMLA) (pp. 559-564).

[15] Liu, S., Rees, B., & Patangia, P. (2024). Supercharging Fraud Detection in Financial Services with Graph Neural Networks | NVIDIA Technical Blog. NVIDIA Technical Blog.

[16] Bespalov, D., Brand, R., & Qi , Y. (2023). Build a GNN-based real-time fraud detection solution using the Deep Graph Library without using external graph storage | Amazon Web Services. Amazon Web Services.

[17] Lu, M., Han, Z., Rao, S. X., Zhang, Z., & Jiang, J. (2022). BRIGHT -- Graph Neural Networks in Real-Time Fraud Detection. https://doi.org/10.48550/arXiv.2205.13084

[18] Motie, S. and Raahemi, B. (2024) 'Financial fraud detection using graph neural networks: A systematic review', *Expert Systems with Applications*, 240, p. 122156. doi:10.1016/j.eswa.2023.122156.

[19] Wang, H., Xu, Z., Li, X., Liu, R., Sheng, Q.Z., Wang, S. and Xie, L., (2021). 'Textual graph neural network for fraud detection'. *IEEE Transactions on Big Data,* 8(5), pp.1255-1266.

[20] Zhang, X., Yao, L., Yuan, F., Atasoy, H., Labrinidis, A. and Vasilakos, A.V. (2023). 'Fraud detection via federated graph neural network'. *IEEE Transactions on Artificial Intelligence*, 4(1), pp.108-121.

[21] Chen, Z., Wang, Y., Zhao, L., Qin, J. and Zhang, D. (2022). 'Federated learning for fraud detection: A comprehensive survey and open challenges. *IEEE Internet of Things Journal*, 9(21), pp.21015-21041.

[22] Li, Y., Wen, J., Li, E., Zhang, Y. and Li, C. (2023). 'A survey on deep learning techniques for credit card fraud detection'. *Information Sciences*, 621, pp.154-177.

[23] Wang, Y., Adams, S., Beling, P., Greenspan, S., Rajagopalan, S., Velez-Rojas, M., ... & Brown, D. (2018). Privacy preserving distributed deep learning and its application in credit card fraud detection. In *2018 17th IEEE Int Conf On Trust, Sec Privacy In Comp Comm / 12th Big Data SE (TrustCom/BigDataSE)* (pp. 1070-1078). IEEE.

[24] Almudaires, F., & Almaiah, M. (2021). Data an overview of cybersecurity threats on credit card companies and credit card risk mitigation. In *2021 Int. Conference on Information Technology (ICIT)* (pp. 732-738). IEEE.

[25] Sakharova, I. (2012). Payment card fraud: Challenges and solutions. In *2012 IEEE international conference on intelligence and security informatics* (pp. 227-234). IEEE.

[26] Btoush, E., Zhou, X., Gururaian, R., Chan, K. C., & Tao, X. (2021). A survey on credit card fraud detection techniques in banking industry for cyber security. In *2021 8th Int Conf on Beh. Social Comp (BESC)* (pp. 1-7). IEEE.

[27] Nuthalapati, A. (2023). Smart Fraud Detection Leveraging Machine Learning For Credit Card Security. *Educational Administration: Theory and Practice*, *29*(2), 433-443.

[28] Motie, S., & Raahemi, B. (2023). Financial fraud detection using graph neural networks: A systematic review. Expert Systems with Applications, 122156. https://doi.org/10.1016/j.eswa.2023.122156

[29] Li, Z., Chen, D., Liu, Q., & Wu, S. (2022). The Devil is in the Conflict: Disentangled Information Graph Neural Networks for Fraud Detection. arXiv (Cornell University). https://doi.org/10.48550/arxiv.2210.12384

[30] Liu, Z., Dou, Y., Yu, P., Deng, Y., & Peng, H. (2020). Virtual Event, China. 4(20). https://doi.org/10.1145/3397271.3401253

[31] Sultana, I., Maheen, S. M., Sunna, A. A., & Kshetri, N. (2024). *SmSeLib: Smart & Secure Libraries-Navigating the Intersection of Machine Learning and Artificial Intelligence - [v1]*. Preprints.org. https://www.preprints.org/manuscript/202411.1445/v1

[32] Kshetri et al (2024). *cryptoRAN: A Review on Cryptojacking and Ransomware Attacks W.R.T. Banking Industry - Threats, Challenges, Problems*. 2 May 2024, ieeexplore.ieee.org/document/10550970, https://doi.org/10.1109/incacct61598.2024.10550970

[33] Shulha, Olha, et al (2022). "Banking Information Resource Cybersecurity System Modeling." *Journal of Open Innovation: Technology, Market, and Complexity*, vol. 8, no. 2, 28 Apr. 2022, p. 80, www.mdpi.com/2199-8531/8/2/80/pdf?version=1651141207, https://doi.org/10.3390/joitmc8020080.

[34] Kshetri, N., Sultana, I., Rahman, M. M., & Shah, D. (2024). DefTesPY: Cyber Defense Model with Enhanced Data Modeling and Analysis for Tesla Company via Python Language. *Ieee.org*, 6. IEEE ETNCC 2024, https://doi.org/10.1109/ETNCC63262.2024.10767532